\documentclass[12pt]{article}
\usepackage{a4}
\usepackage{bm}
\usepackage{graphicx}

\usepackage{amsmath}
\usepackage{amsfonts}

\newcommand{\CC}{\mathbb{C}}
\newcommand{\ZZ}{\mathbb{Z}}
\newcommand{\RR}{\mathbb{R}}

\newcommand{\tr}{{\rm tr}}
\newcommand{\ol}{\overline}
\newcommand{\CP}{\CC\bm{P}}

\def\rank{\mathop{\rm rank}\nolimits}
\def\sign{\mathop{\rm sign}\nolimits}
\def\diag{\mathop{\rm diag}\nolimits}
\def\PE{\mathop{\rm PE}\nolimits}

\begin{document}
\begin{titlepage}
\title{
\begin{flushright}
\normalsize{TIT/HEP-622\\
Oct 2012}
\end{flushright}
       \vspace{2cm}
Perturbative partition function for a squashed ${\bm S}^5$
       \vspace{2cm}}
\author{
Yosuke Imamura\thanks{E-mail: \tt imamura@phys.titech.ac.jp}$^{~1}$
\\[30pt]
{\it $^1$ Department of Physics, Tokyo Institute of Technology,}\\
{\it Tokyo 152-8551, Japan}
}
\date{}

\maketitle
\thispagestyle{empty}

\vspace{0cm}

\begin{abstract}
We compute the index of 6d ${\cal N}=(1,0)$ theories
on $\bm S^5\times\RR$ containing
vector and hypermultiplets.
We only consider the perturbative sector without instantons.
By compactifying $\RR$ to $\bm S^1$ with a twisted boundary condition
and taking the small radius limit,
we derive the perturbative partition function on a squashed ${\bm S}^5$.
The $1$-loop partition function is represented in a simple form
with the triple sine function.
\end{abstract}

\end{titlepage}

\section{Introduction}
The sphere partition functions play important roles in the recent progress
in supersymmetric gauge theories and string/M-theory.
The ${\bm S}^4$ partition function \cite{Pestun:2007rz}
provides data on the 4d side of the AGT relation \cite{Alday:2009aq},
which connects 4d ${\cal N}=2$ supersymmetric gauge theories and
2d conformal field theories.
The ${\bm S}^3$ partition function \cite{Kapustin:2009kz,Jafferis:2010un,Hama:2010av} is used to investigate the dynamics
of 3d supersymmetric gauge theories.
In particular, it
reproduces the $N^{3/2}$ behavior of the free energy
of 3d gauge theories
realized on M2-branes \cite{Drukker:2010nc,Herzog:2010hf,Marino:2011eh}.
The ${\bm S}^5$ partition function \cite{Kallen:2012cs,Kallen:2012va,Kim:2012av}
also attracts great interest recently.
The motivation comes from
interests in 5d conformal field theories \cite{Seiberg:1996bd}
and
the conjecture \cite{Douglas:2010iu,Lambert:2010iw}
concerning the relation
between 5d supersymmetric Yang-Mills theories
and mysterious 6d $(2,0)$ supersymmetric theories.
See also \cite{Hosomichi:2012ek,Kawano:2012up,Kim:2012gu,Terashima:2012ra,Kim:2012tr,Fukuda:2012jr}
for resent studies of 5d supersymmetric Yang-Mills theories on various backgrounds.

These partition functions depend on continuous parameters.
This fact is important because
we can extract physical information from the
dependence on the continuous parameters.
Mass parameters, such as real mass parameters
in 3d ${\cal N}=2$ gauge theories and complex ones
in 4d ${\cal N}=2$ gauge theories can be easily turned on as
expectation values of background vector multiplets.
We also have a similar mass parameter
in 5d ${\cal N}=2$ case \cite{Kim:2012av}.
The introduction of deformation parameters
of spheres is more involved.
Two kinds of deformed spheres have been studied in the literature:
ellipsoids and squashed spheres.\footnote{The term ``squashed spheres'' often
refers to both kinds of deformations
in the literature.
In this paper, for distinction, we use this term only for the
specific deformation of the sphere which is presented in \S\ref{squashed.sec}.}
The partition function for 3d ellipsoids
is computed in \cite{Hama:2011ea},
and the same form of the partition function is reproduced
for squashed $\bm S^3$ in \cite{Imamura:2011wg}.
The partition function for  4d ellipsoids
is worked out in \cite{Hama:2012bg}
and the consistency to the AGT relation is found.
The purpose of this paper is to determine the
partition function for squashed ${\bm S}^5$.

A convenient way to construct a squashed $\bm S^n$ is
dimensional reduction from $\bm S^n\times\RR$.
If we compactify $\RR$ with a twisted boundary condition
we obtain a squashed $\bm S^n$.
Supersymmetric theories on squashed $\bm S^3$ are constructed in
\cite{Imamura:2011wg}
by using this dimensional reduction,
and their one-loop partition function can be obtained
by taking a certain limit of the 4d index \cite{Dolan:2011rp,Gadde:2011ia,Imamura:2011uw}.
In this paper we apply the same method to
5d ${\cal N}=1$ supersymmetric gauge theories
and derive the partition function in the perturbative sector
for squashed $\bm S^5$ with general squashing parameters $\phi_i$ ($i=1,2,3$).
The final result is
\begin{equation}
Z_{\rm pert}
=C({\bm\omega})\left(\prod_{i=1}^{\rank G}\int_{-\infty}^\infty d\sigma_i\right)\exp\left[-\frac{(2\pi)^3}{\omega_1\omega_2\omega_3}{\cal F}(\sigma)\right]
\frac
{
\prod_{\alpha\in{\rm root}}
S_3(-i\alpha(\sigma),{\bm\omega})
}
{
\prod_{\rho\in R}
S_3(-i\rho(\sigma)+\frac{\omega_1+\omega_2+\omega_3}{2},{\bm\omega})
},
\label{final}
\end{equation}
where ${\cal F}$ is the prepotential of the theory
and
$S_3(z,\bm\omega)$ is the triple sine function with
periods $\bm\omega=(\omega_1,\omega_2,\omega_3)$.
The squashing parameters appear in this
partition function through the periods $\omega_i=1+i\phi_i$.
We assume that the hypermultiplets belong to the representation $R+\ol R$
of the gauge group.
$\alpha$ runs over all (positive and negative) roots of the gauge group $G$,
and $\rho$ runs over the weights in the representation $R$.
We can introduce mass parameters for hypermultiplets straightforwardly by
shifting the weights $\rho$.
The prepotential ${\cal F}$ is normalized so that
the Chern-Simons term is given by
\begin{equation}
S=\frac{i}{6}\frac{\partial^3{\cal F}(\sigma)}{\partial\sigma^\alpha\partial\sigma^\beta\partial\sigma^\gamma}\int A^\alpha\wedge F^\beta\wedge F^\gamma+\cdots.
\label{chser}
\end{equation}
The overall factor $C({\bm\omega})$ depends only on ${\bm\omega}$,
and its explicit form is given in (\ref{comega}).

This paper is organized as follows.
In the next section, we summarize the background manifold that we consider and
the supersymmetry on it.
In \S\ref{index.sec} we compute the index of
vector and hypermultiplets in 6d ${\cal N}=(1,0)$ theories by localization.
In section \ref{limit.sec},
we take the small radius limit of the index and obtain
the $\bm S^5$ partition function.
The last section is devoted to discussion.

\subsection*{Note added}
While this work was being completed,
there appeared a paper by
G.~Lockhart and C.~Vafa \cite{LockhartVafa},
which also studies the partition function for a deformed $\bm S^5$.

\section{Background manifold and supersymmetry}
\subsection{Squashed sphere}\label{squashed.sec}
A squashed $\bm S^{2r-1}$ is easily obtained by
the dimensional reduction from $2r$-dimensional manifold
$\bm S^{2r-1}\times\RR$ with the metric
\begin{equation}
ds^2=dt^2+\sum_{i=1}^r|dz_i|^2.
\end{equation}
In this paper we only consider the $r=3$ case.
The complex coordinates $z_i$ ($i=1,\ldots,r$) are constrained by
\begin{equation}
\sum_{i=1}^r|z_i|^2=R^2.
\label{zz1}
\end{equation}
In the following we set $R=1$.
If we compactified the ``time'' direction by
the identification $(t,z_i)\sim(t+\beta,z_i)$
and took the small radius limit $\beta\rightarrow 0$,
we would obtain the round ${\bm S}^{2r-1}$.
Instead, we consider the twisted identification
\begin{equation}
(t,z_i)\sim(t+\beta,e^{i\phi_i\beta}z_i).
\label{ident}
\end{equation}
The parameters $\phi_i$ are called squashing parameters.
If we introduce real coordinates $\rho_i$ and $\theta_i$ by
\begin{equation}
\rho_ie^{i\theta_i}=e^{-i\phi_it}z_i,
\end{equation}
the identification becomes
$(t,\rho_i,\theta_i)\sim(t+\beta,\rho_i,\theta_i)$.
In the new coordinate system
the metric becomes
\begin{equation}
ds_6^2=v^2(dt+W)^2+ds_5^2,
\label{ds6vds5}
\end{equation}
where $v$ and $W$ are the scalar function and the differential
defined by
\begin{equation}
v^2=1+\sum_{i=1}^r \phi_i^2\rho_i^2,\quad
W=\frac{1}{v^2}\sum_{i=1}^r\phi_i\rho_i^2d\theta_i.
\end{equation}
$ds_5^2$
in (\ref{ds6vds5})
is
the metric of the squashed sphere.
Its explicit form is
\begin{equation}
ds_5^2=\sum_{i=1}^r(d\rho_i^2+\rho_i^2d\theta_i^2)-v^2W^2.
\end{equation}

\subsection{${\cal N}=(1,0)$ supersymmetry on ${\bm S}^5\times\RR$}
The 6d ${\cal N}=(1,0)$ superconformal symmetry
on an arbitrary conformally flat background
is parameterized by symplectic
Majorana-Weyl spinors $\epsilon$ and $\kappa$
satisfying
\begin{equation}
D_M\epsilon=\Gamma_M\kappa.
\label{6dkilling}
\end{equation}
The spinors
$\epsilon$ and $\kappa$ have positive and negative chirality,
respectively.
We use $M, N,\ldots$ for 6d vector indices
and $\mu,\nu,\ldots$ for 5d ones.
6d and 5d Dirac matrices are denoted by $\Gamma_M$ and $\gamma_\mu$, respectively.
Our choice of the representation of Dirac matrices are given in Appendix.
The R-symmetry
is $SU(2)_R$, and
both $\epsilon$ and $\kappa$ are $SU(2)_R$ doublets.
We use $I,J,\ldots=1,2$ for $SU(2)_R$ indices when we show them explicitly.

Let us consider supersymmetry on $\bm S^5\times \RR$.
We use vector indices $\mu,\nu,\ldots=1,\ldots,5$ for $\bm S^5$
and $6$ for $\RR$.
The ${\cal N}=(1,0)$ supersymmetry algebra contains
$16$ supercharges.
Because ${\bm S}^5\times\RR$ is conformally flat,
we can realize all $16$ supersymmetries on ${\bm S}^5\times\RR$ as well
as on $\RR^6$,
and the Killing equation (\ref{6dkilling})
has $16$ linearly independent solutions.
The general solution is
\begin{equation}
\epsilon_I=\left(\begin{array}{c}
e^{-\frac{t}{2}}\varepsilon_i({\bm4})\eta_I^i
+e^{+\frac{t}{2}}\varepsilon^i(\ol{\bm4})\ol\eta_{Ii} \\
0
\end{array}\right),
\end{equation}
where $\eta_I^i$ and $\ol\eta_{Ii}$ ($i=1,2,3,4$)
are Grassmann-odd constant parameters,
and
$\varepsilon_i({\bm4})$ and
$\varepsilon^i(\ol{\bm4})$ are
basis of Killing spinors on ${\bm S}^5$
satisfying
\begin{equation}
D_\mu\varepsilon_i({\bm4})=-\frac{i}{2r}\gamma_\mu\varepsilon_i({\bm 4}),\quad
D_\mu\varepsilon^i(\ol{\bm4})=+\frac{i}{2r}\gamma_\mu\varepsilon^i(\ol{\bm 4}).
\end{equation}
Note that $\epsilon_I$ are eight-component 6d spinors while $\varepsilon_i(\bm4)$ and
$\varepsilon^i(\ol{\bm4})$ are
defined as
four-component 5d spinors.
Let $SO(6)_{\rm iso}$ be the isometry of $\bm S^5$.
$\varepsilon_i({\bm4})$ and
$\varepsilon^i(\ol{\bm4})$ belong to
${\bm4}$ and $\ol{\bm4}$ of $SO(6)_{\rm iso}$, respectively.
$16$ parameters $\eta_I^i$ and $\ol\eta_{Ii}$ correspond
to the $16$ supercharges in the ${\cal N}=(1,0)$ superconformal algebra.
The spinors $\epsilon$ and $\kappa$ are related by
\begin{equation}
\kappa=-\frac{1}{2r}\Gamma_{\rm iso}\Gamma^6\epsilon,
\end{equation}
where
$\Gamma_{\rm iso}$ is the
$SO(6)_{\rm iso}$ chirality operator
that acts on
$\varepsilon_i({\bm 4})$ and $\varepsilon^i(\ol{\bm4}$)
as $+1$ and $-1$, respectively.

We consider a gauge theory
whose action contains the Yang-Mills term.
Because $1/g_{\rm YM}^2$ has mass dimension one in 5d and
the Yang-Mills term is a kind of mass deformation,
the superconformal symmetry in 5d is broken to the rigid supersymmetry.
Correspondingly, we impose the condition
\begin{equation}
\Gamma_{\rm iso}\epsilon=\tau_3\epsilon
\label{rigidcond}
\end{equation}
on the parameter $\epsilon$.
This condition admits eight supersymmetries with parameters
\begin{equation}
\eta_1^i\equiv \eta^i,\quad
\eta_2^i=0,\quad
\ol\eta_{1i}=0,\quad
\ol\eta_{2i}\equiv\ol\eta_i.
\end{equation}
The general Killing spinor for rigid supersymmetry has components
\begin{equation}
\epsilon_1=\left(\begin{array}{c}
e^{-\frac{t}{2}}\varepsilon_i({\bm4})\eta^i \\ 0
\end{array}\right),\quad
\epsilon_2=\left(\begin{array}{c}
e^{+\frac{t}{2}}\varepsilon^i(\ol{\bm4})\ol\eta_i \\ 0
\end{array}\right).
\end{equation}

For the computation of the index
with the help of localization technique,
we have to choose one supersymmetry $\cal Q$.
Before giving our choice of ${\cal Q}$
we first choose a complex structure (K\"ahler form)
$I_{\CC^3}$ on $\CC^3$ spanned by $z_i$ in (\ref{zz1}),
\begin{equation}
I_{\CC^3}=-\frac{i}{2}dz_i^*\wedge dz_i.
\label{ic3}
\end{equation}
Let $SU(3)_V\times U(1)_V$ be the subgroup
of $SO(6)_{\rm iso}$ which keeps (\ref{ic3}) invariant.
$U(1)_V$ is generated by $I_{\CC^3}$, and we parameterize
$U(1)_V$-orbits by a coordinate $\psi$ so that
\begin{equation}
I_{\CC^3}=-{\cal L}_\psi,
\end{equation}
where ${\cal L}_\psi$ is the Lie derivative along the vector $\partial_\psi$.
With this $U(1)_V$ symmetry we naturally represent
${\bm S}^5$ as the Hopf fibration over $\CP^2$.
$I_{\CC^3}$ naturally induces
the complex structure (K\"ahler form) of $\CP^2$,
and its pullback to $\bm S^5\times\RR$
is denoted by $I$.

The metric on ${\bm S}^5\times\RR$ is
\begin{equation}
ds^2=e^Me^M
=e^me^m+e^5e^5+e^6e^6,\quad
(M=1,\ldots,6,\quad
m=1,2,3,4),
\end{equation}
where
$e^5$ and $e^6$ are
\begin{equation}
e^5=d\psi+V,\quad
e^6=dt.
\end{equation}
$V$ is a differential on $\CP^2$ satisfying $dV=-2I$,
and $t\equiv x^6$ is the coordinate along $\RR$.
We choose the vielbein on $\CP^2$, $e^m$ ($m=1,2,3,4$), so that
\begin{equation}
I=e^2\wedge e^1+e^4\wedge e^3.
\end{equation}

The choice of the complex structure $I_{\CC^3}$ breaks
$SO(6)_{\rm iso}$ to $SU(3)_V\times U(1)_V$,
and the representations $\bm4$ and $\ol{\bm4}$ branch into
$SU(3)_V\times U(1)_V$ irreducible representations
as
\begin{equation}
{\bm 4}={\bm 3}_{+\frac{1}{2}}+{\bm 1}_{-\frac{3}{2}},\quad
\ol{\bm 4}=\ol{\bm 3}_{-\frac{1}{2}}+{\bm 1}_{+\frac{3}{2}}.
\label{branching0}
\end{equation}
Correspondingly, Killing spinors split as
\begin{equation}
\varepsilon_i({\bm4})
\rightarrow\left(
\varepsilon_i({\bm3}_{+\frac{1}{2}}),
\varepsilon({\bm1}_{-\frac{3}{2}})
\right),\quad
\varepsilon^i(\ol{\bm4})
\rightarrow\left(
\varepsilon^i(\ol{\bm3}_{-\frac{1}{2}}),
\varepsilon({\bm1}_{+\frac{3}{2}})
\right).
\end{equation}
We introduce the shorthand notation for the $SU(3)_V$ singlet Killing spinors
\begin{equation}
\xi_1=\varepsilon({\bm1}_{-\frac{3}{2}}),\quad
\xi_2=\varepsilon({\bm1}_{+\frac{3}{2}}).
\end{equation}
With our choice of the Dirac matrices given in Appendix,
$\xi_1$ and $\xi_2$ have the components
\begin{equation}
\xi_1=(0,1,0,0)^T,\quad
\xi_2=(1,0,0,0)^T,
\end{equation}
and they are related to $I$ by
\begin{equation}
I_{\mu\nu}
=-i\xi_1^\dagger\gamma_{\mu\nu}\xi_1
=i\xi_2^\dagger\gamma_{\mu\nu}\xi_2.
\end{equation}

Now we are ready to give our choice of ${\cal Q}$.
We define ${\cal Q}$ by
\begin{equation}
\eta{\cal Q}=\delta(\eta\varepsilon),
\end{equation}
where the right hand side is the SUSY transformation
with the parameter $\epsilon=\eta\varepsilon$.
$\eta$ is a Grassmann-odd constant
and $\varepsilon$ is the 6d Killing spinor
with components
\begin{equation}
\varepsilon_1=\left(\begin{array}{c}
e^{-\frac{t}{2}}\xi_1 \\
0
\end{array}\right),\quad
\varepsilon_2=\left(\begin{array}{c}
-e^{+\frac{t}{2}}\xi_2 \\
0
\end{array}\right).
\label{thevareps}
\end{equation}

\section{Index of $(1,0)$ theory}\label{index.sec}
\subsection{Definition of the index}
In this section, we derive the 5d partition function as a
limit of the index of the corresponding 6d ${\cal N}=(1,0)$ theory.
To obtain a 5d theory with a gauge group $G$ and the matter representation $R$,
we start from the 6d theory with the same $G$ and $R$.
Although the 5d theory is well-defined, its 6d counterpart may be anomalous.
Even so, we can still use the ``index''
as a concise way to express the spectrum of
fluctuations
of fields on ${\bm S}^5\times{\bm S}^1$
around saddle points.
We are only interested in the modes that survive after the small radius limit
$\beta\rightarrow 0$,
and
the ``index'' is useful to obtain such modes
although it cannot be regarded as a physical quantity when
the theory is anomalous.

The bosonic symmetry of the 6d ${\cal N}=(1,0)$ theory on $\bm S^5 \times\RR$
is $\RR\times SO(6)_{\rm iso}\times SU(2)_R$,
and the choice of ${\cal Q}$ breaks it to
$\RR\times U(1)_V\times SU(3)_V\times SU(2)_R$.
Correspondingly, there are five Cartan generators
\begin{equation}
H=-\partial_t,\quad
Q_V=-i{\cal L}_\psi,\quad
\tau_3,\quad
\lambda_3,\quad
\lambda_8.
\label{cartangen}
\end{equation}
$\tau_3$ is the $SU(2)_R$ Cartan generator acting on doublets as
Pauli matrix $\tau_3=\diag(1,-1)$.
$\lambda_3$ and $\lambda_8$ are
$SU(3)_V$ Cartan generators whose fundamental representation
matrices are
\begin{equation}
\lambda_3|_{\bm3}=\left(\begin{array}{ccc}
1 \\
& -1 \\
&& 0
\end{array}\right),\quad
\lambda_8|_{\bm3}=\left(\begin{array}{ccc}
1 \\
& 1 \\
&& -2
\end{array}\right).
\end{equation}
The quantum numbers of $\varepsilon_I$ in
(\ref{thevareps})
are shown in Table
\ref{table:varep}.
\begin{table}[htb]
\caption{Quantum numbers of $\varepsilon_I$.}
\label{table:varep}
\begin{center}
\begin{tabular}{cccccc}
\hline
\hline
& $H$ & $Q_V$ & $\tau_3$ & $\lambda_3$ & $\lambda_8$ \\
\hline
$\varepsilon_1$ & $+\frac{1}{2}$ & $-\frac{3}{2}$ & $+1$ & $0$ & $0$ \\
$\varepsilon_2$ & $-\frac{1}{2}$ & $+\frac{3}{2}$ & $-1$ & $0$ & $0$ \\
\hline
\end{tabular}
\end{center}
\end{table}
The general transformation generated by (\ref{cartangen})
that commutes with the supersymmetry ${\cal Q}$
is
\begin{equation}
{\cal O}=
q^{H-Q_V-2\tau_3}x^{Q_V+\frac{3}{2}\tau_3}y_3^{\lambda_3}y_8^{\lambda_8}.
\label{theopo}
\end{equation}
The index is defined with this operator by \cite{Bhattacharya:2008zy}
\begin{equation}
{\cal I}(x,y_3,y_8)=\tr\left[
(-1)^F
{\cal O}
\right],
\label{indexdef}
\end{equation}
where $F$ is the fermion number operator
and the trace is taken over the
Fock space of gauge invariant states
on ${\bm S}^5$.
This can be computed by the path integral of the
theory defined on ${\bm S}^5\times{\bm S}^1$.
The path integral in the perturbative sector, which does not contain
instantons on $\CP^2$,
reduces to the finite-dimensional integral
\begin{equation}
{\cal I}=\oint[d\sigma]Z_0\PE{\cal I}_{\rm sp}.
\end{equation}
$Z_0$ is the zero-point contribution and
${\cal I}_{\rm sp}$ is the single-particle index defined by
(\ref{indexdef}) with the trace replaced by the summation
over single-particle excitations
including charged particles.
$\PE$ is the plethystic exponential defined by
\begin{equation}
\PE f(x_i)=\exp\sum_{m=1}^\infty\frac{1}{m}f(x_i^m).
\end{equation}
In general, the single-particle index takes the form
\begin{equation}
{\cal I}_{\rm sp}=\sum_i n_iq^{a_i}x^{b_i}y_3^{c_i}y_8^{d_i},
\end{equation}
and then
the corresponding zero-point contribution $Z_0$
and the plethystic exponential $\PE{\cal I}_{\rm sp}$
are
\begin{eqnarray}
Z_0&=&\prod_i
q^{\frac{n_i}{2}a_i}
x^{\frac{n_i}{2}b_i}
y_3^{\frac{n_i}{2}c_i}
y_8^{\frac{n_i}{2}d_i},\nonumber\\
\PE{\cal I}_{\rm sp}&=&\prod_i\left(1-q^{a_i}x^{b_i}y_3^{c_i}y_8^{d_i}\right)^{-n_i}.
\end{eqnarray}
We define ``modified plethystic exponential'' as
the product of these two factors, i.e.,
\begin{equation}
\PE'{\cal I}_{\rm sp}
\equiv Z_0\PE{\cal I}_{\rm sp}
=\prod_i\left(
q^{\frac{a_i}{2}}x^{\frac{b_i}{2}}y_3^{\frac{c_i}{2}}y_8^{\frac{d_i}{2}}
-q^{-\frac{a_i}{2}}x^{-\frac{b_i}{2}}y_3^{-\frac{c_i}{2}}y_8^{-\frac{d_i}{2}}
\right)^{-n_i}.
\end{equation}
The integration variable
$\sigma$ is the $t$-component of the gauge field (See Eq. (\ref{bkggauge}))
and the integration over $\sigma$ picks up
gauge invariant states.
The integration measure is
\begin{equation}
\oint[d\sigma]
=C_0\prod_{i=1}^{\rank G}\left(
\oint_0^{2\pi/\beta} d\sigma_i\right) J(\sigma),\quad
J(\sigma)\equiv \prod_{\alpha\in{\rm root}}2i \sin\frac{\beta\alpha(\sigma)}{2}.
\label{intmeasure}
\end{equation}
Each diagonal component of $\sigma$ takes value in ${\bm S}^1$ with period $2\pi/\beta$.
The factor $C_0$ is given by
\begin{equation}
C_0=\frac{1}{|W|}\left(\frac{\beta}{2\pi}\right)^{\rank G},
\end{equation}
where $|W|$ is the order of the Weyl group of $G$.
This factor is needed to guarantee that each gauge invariant
state contributes
to the index with the correct weight.

\subsection{Vector multiplet}
The SUSY transformation laws of
${\cal N}=(1,0)$ vector multiplets are
\begin{eqnarray}
\delta A_M&=&-(\epsilon\Gamma_M\lambda),\nonumber\\
\delta\lambda&=&-\frac{1}{2}\Gamma^{MN}F_{MN}\epsilon+iD_a\tau_a\epsilon,\nonumber\\
\delta D_a&=&i(\epsilon\tau_a\Gamma^M D_M\lambda)-2i(\kappa\tau_a\lambda),
\end{eqnarray}
where $\tau_a$ ($a=1,2,3$) are $SU(2)_R$ generators.
To localize the path integral,
we need to deform the action by ${\cal Q}$-exact terms.
A standard choice of the ${\cal Q}$-exact term is
\begin{equation}
{\cal L}\propto {\cal Q}[({\cal Q}\lambda)^\dagger\lambda],
\label{lusual}
\end{equation}
where ``$\dagger$'' is the Hermitian conjugate.
Although the bosonic part of this Lagrangian is positive semi-definite,
this action is not suitable for computation of the index
because this has unwanted explicit $t$-dependence.
For example,
the norm $\varepsilon^\dagger\varepsilon$
of the Killing spinor $\varepsilon$
defined by (\ref{thevareps})
appears in (\ref{lusual}), and
depends on the coordinate $t$;
\begin{equation}
\varepsilon^\dagger\varepsilon=2\cosh t.
\end{equation}
To avoid such $t$-dependence,
we introduce the new conjugate operator ``$\star$''
that acts on $\varepsilon$ as
\begin{equation}
\varepsilon^\star=\varepsilon^\dagger|_{t\rightarrow -t},
\end{equation}
and on all other bosonic fields as ``$\dagger$''.
The norm of $\varepsilon$ defined with $\star$
is constant;
\begin{equation}
\varepsilon^\star\varepsilon=2.
\end{equation}
We use the following ${\cal Q}$-exact terms defined with
$\star$, which has no explicit $t$-dependence.
\begin{eqnarray}
&&\frac{s}{2}{\cal Q}[({\cal Q}\lambda)^\star \lambda]
\nonumber\\
&&=s\bigg[\frac{1}{2}F_{MN}F^{MN}-\frac{1}{4}\epsilon^{mnpq56}F_{mn}F_{pq}
-D_aD_a
\nonumber\\
&&
-\lambda\Gamma^M D_M\lambda
-\frac{i}{4}I_{MN}(\lambda\Gamma^{MN}\Gamma_6\lambda)
-\frac{i}{2}(\lambda\Gamma_5\tau_3\lambda)
+\frac{1}{2}(\lambda\Gamma_6\tau_3\lambda)\bigg].
\label{vqexact}
\end{eqnarray}
The path integral does not depend on the deformation parameter $s\in\RR_+$,
and in the weak coupling limit $s\rightarrow\infty$
it reduces to the Gaussian integral.
The path integral of the auxiliary fields $D_a$ gives
constant and we neglect it.

The saddle point equations for the vector field are
\begin{equation}
F_{12}+F_{34}=
F_{23}+F_{14}=
F_{31}+F_{24}=
F_{5m}=
F_{6m}=
F_{56}=0.
\label{saddlept}
\end{equation}
In the perturbative sector with vanishing instanton number on $\CP^2$,
(\ref{saddlept}) implies $F_{MN}=0$,
and up to gauge transformations
we can set the gauge potential to be
\begin{equation}
A=\sigma dt,
\label{bkggauge}
\end{equation}
where $\sigma$ is a constant diagonal matrix.
The path integral of $A_M$ and $\lambda$ around these saddle points
can be explicitly performed in the weak coupling limit.

Let us first consider the bosonic sector.
The bosonic part of the ${\cal Q}$ exact action,
together with the gauge fixing term $(D_\mu A^\mu)^2$,
is
\begin{equation}
{\cal L}_A
=s\left(
\frac{1}{2}F_{MN}F^{MN}-\frac{1}{4}\epsilon^{mnpq56}F_{mn}F_{pq}
+(D_\mu A_\mu)^2\right).
\label{bosonica}
\end{equation}
To perform the path integral,
we expand fluctuations of $A_M$ by harmonics on $\CP^2$.
See Refs. \cite{Kim:2012av,Pope:1980ub} for detailed explanation
for the harmonic expansion.
The harmonics are classified by
$SU(3)_V$ representations
labeled by two quantum numbers $k$ and $m$.
They are related to the
$SU(3)$ Dynkin labels $k_1,k_2\in\ZZ_{\geq0}$ by
\begin{equation}
k=k_1+k_2,\quad
m=k_2-k_1.
\end{equation}
For example, $(k,m)=(1,-1)$ and $(k,m)=(1,1)$
correspond to the fundamental and anti-fundamental representations, respectively.
Because both $k_1$ and $k_2$ are non-negative integers,
$m$ for specific $k$ runs
from $-k$ to $k$ with step $2$, i.e.,
\begin{equation}
m=-k,-k+2,\ldots,k-2,k.
\end{equation}

Let $\mu$ be
an $SU(3)_V$ weight vector
and $\alpha$ be a weight vector in the adjoint representation
of the gauge group.
For specific $\mu$ and $\alpha$,
there are generically six bosonic modes.
Let $a_i$ ($i=1,\ldots,6$) be the
corresponding coefficients in the harmonic expansion of $A_M$.
By substituting the mode expansion to the action 
(\ref{bosonica}) we obtain the quadratic form $a_iM_{ij}a_j$
with the matrix $M_{ij}$ depending on the quantum numbers $k$, $m$,
and the Wilson line $\sigma$.
The determinant of $M_{ij}$ is
\begin{equation}
[k(k+4)]^2
(k^2-D_6^2)
[(k+4)^2-D_6^2]
[k(k+4)-2m+9-D_6^2]
[k(k+4)+2m+9-D_6^2],
\label{bosvdet}
\end{equation}
up to an unimportant constant.
The covariant derivative $D_6$ in
(\ref{bosvdet}) contains the background gauge field (\ref{bkggauge});
\begin{equation}
D_6
=\partial_6-i[\sigma,*]
=\partial_6-i\alpha(\sigma).
\end{equation}

The factor $[k(k+4)]^2$ is canceled by the Faddeev-Popov determinant
associated with the gauge fixing term $(D_\mu A_\mu)^2$,
and the other factors represent the contribution of
physical modes.
We can read off one-particle excitations
on ${\bm S}^5$ as zeros of (\ref{bosvdet}).
We show the spectrum obtained in this way in Table \ref{table:vector}.
\begin{table}[htb]
\caption{Single particle states of the vector field $A_M$.
$\sqrt\pm$ represents $\sqrt{k(k+4)\pm 2m+9}$ and $\alpha$ stands for $\alpha(\sigma)$.}
\label{table:vector}
\begin{center}
\begin{tabular}{clccclc}
\hline
\hline
ID & $H$  & $Q_V$ & $\tau_3$ & Range of $m$ & $H-Q_V-2\tau_3$ & $Q+\frac{3}{2}\tau_3$ \\
\hline
$[A1]$ & $-i\alpha+ k$ & $m$ & $0$ & $-k+2\leq m\leq k-2$ & $-i\alpha+k-m$ & $m$ \\
$[A2]$ & $-i\alpha+k+4$ & $m$ & $0$  & $-k\leq m\leq k$ & $-i\alpha+k-m+4$ & $m$ \\
$[A3]$ & $-i\alpha+\sqrt{+}$ & $m+3$ & $0$  & $-k\leq m\leq k-2$ & $-i\alpha+\sqrt{+}-m-3$ & $m+3$ \\
$[A4]$ & $-i\alpha+\sqrt{-}$ & $m-3$ & $0$  & $-k+2\leq m\leq k$ & $-i\alpha+\sqrt{-}-m+3$ & $m-3$ \\
\hline
\end{tabular}
\end{center}
\end{table}

If $m=\pm k$, some modes in the table are absent.
Furthermore,
special treatment is needed when $k=0$.
In this case, in addition to modes shown in Table \ref{table:vector},
we have a bosonic zero mode of $A_6$
corresponding to a residual gauge symmetry.
The gauge fixing term $\propto(D_\mu A_\mu)$
in (\ref{bosonica})
does not fix gauge transformations
with the parameter depending only on $x^6$.
We impose further gauge fixing condition
which makes $A_6$ be diagonal constant matrix
as in (\ref{bkggauge}).
The integration measure $J(\sigma)$
in (\ref{intmeasure})
is the Faddeev-Popov determinant associated with this
extra gauge fixing.

Next, let us consider the fermionic sector.
The $\lambda$-bilinear part in the $\cal Q$-exact action
(\ref{vqexact})
is
\begin{equation}
{\cal L}_\lambda=s\left[
-\lambda\left(
\Gamma^MD_M-\frac{1}{4}I_{MN}\Gamma^{MN}+\frac{i}{2}\tau_3\Gamma^5-\frac{i}{2}\tau_3\right)\lambda\right].
\label{fermionbin}
\end{equation}
By expanding $\lambda$ by spinor harmonics on $\CP^2$,
we obtain the spectrum shown in Table
\ref{gauginospec.tbl}.
\begin{table}[htb]
\caption{One particle spectrum of $\lambda$.
$\sqrt{\pm}$ and $\alpha$ stand for $\sqrt{k(k+4)\pm 2m+9}$ and $\alpha(\sigma)$, respectively.}
\label{gauginospec.tbl}
\begin{center}
\begin{tabular}{clccclc}
\hline
\hline
ID & $H$ & $Q_V$ & $\tau_3$ & Range of $m$ & $H-Q_V-2\tau_3$ & $Q_V+\frac{3}{2}\tau_3$ \\
\hline
$[\lambda 1]$ & 
$-i\alpha+k+\frac{1}{2}$ & $m-\frac{3}{2}$ & $+1$ & $-k+2\leq m\leq k$ & $-i\alpha+k-m$ & $m$ \\
$[\lambda 2]$ & 
$-i\alpha+k+\frac{7}{2}$ & $m+\frac{3}{2}$ & $-1$ & $-k\leq m\leq k$ & $-i\alpha+k-m+4$ & $m$ \\
$[\lambda 3]$ & 
$-i\alpha+\sqrt{+}+\frac{1}{2}$ & $m+\frac{3}{2}$ &$+1$& $-k\leq m\leq k$ & $-i\alpha+\sqrt{+}-m-3$ & $m+3$ \\
$[\lambda 4]$ & 
$-i\alpha+\sqrt{-}-\frac{1}{2}$ & $m-\frac{3}{2}$ &$-1$& $-k+2\leq m\leq k$ & $-i\alpha+\sqrt{-}-m+3$ & $m-3$ \\
\hline
\end{tabular}
\end{center}
\end{table}

Let us compare the bosonic spectrum and the fermionic one.
The bosonic modes $[A1]$, $[A2]$, $[A3]$, and $[A4]$
have the same quantum numbers
$H-Q_V-2\tau_3$ and $Q_V+\frac{3}{2}\tau_3$
as the fermionic modes
$[\lambda1]$, $[\lambda2]$, $[\lambda3]$, and $[\lambda4]$, respectively,
and almost all of them cancel to each other.
Only
the fermionic modes $[\lambda1]$ and $[\lambda3]$ with $m=k$ are left
due to the difference in the range of $m$.
(Note that $k-m$ must be an even integer and there are no $m=k-1$ modes.)
These surviving modes contribute to the
single-particle index by
\begin{equation}
-\sum_{\alpha\in{\rm adj}}
\left[
\sum_{k=1}^\infty q^{-i\alpha(\sigma)}x^k\chi_{(k,k)}(y_3,y_8)
+\sum_{k=0}^\infty q^{-i\alpha(\sigma)}x^{k+3}\chi_{(k,k)}(y_3,y_8)
\right],
\label{ispp0}
\end{equation}
where $\chi_{(k,m)}(y_3,y_8)$ is the
$SU(3)_V$ character
\begin{equation}
\chi_{(k,m)}=\sum_{\mu\in(k,m)}y_3^{\mu(\lambda_3)}y_8^{\mu(\lambda_8)}.
\end{equation}
Note that $\alpha$ in
(\ref{ispp0}) runs over all weights in the adjoint representation of
the gauge group.
Let us separate the Cartan part
\begin{equation}
{\cal I}_{\rm sp}^{\rm Cartan}
=-(\rank G)
\left[
\sum_{k=1}^\infty x^k\chi_{(k,k)}(y_3,y_8)
+\sum_{k=0}^\infty x^{k+3}\chi_{(k,k)}(y_3,y_8)
\right],
\end{equation}
and combine the remaining part
with
$-\sum_{\alpha\in{\rm root}} q^{-i\alpha}$,
which reproduces the measure factor
\begin{equation}
J(\sigma)=\PE'\left(-\sum_{\alpha\in{\rm root}}q^{-i\alpha(\sigma)}\right).
\end{equation}
Then we have the single particle index
\begin{equation}
{\cal I}^{\rm vector}_{\rm sp}(q,x,y_3,y_8)
=-
\sum_{\alpha\in{\rm root}}
\sum_{k=0}^\infty
\left[
q^{-i\alpha(\sigma)}x^k\chi_{(k,k)}(y_3,y_8)
+q^{-i\alpha(\sigma)}x^{k+3}\chi_{(k,k)}(y_3,y_8)
\right].
\label{ispp}
\end{equation}
With this definition of ${\cal I}^{\rm vector}_{\rm sp}$,
we do not have to include $J(\sigma)$ separately in (\ref{intmeasure}).

\begin{figure}[htb]
\centerline{\includegraphics{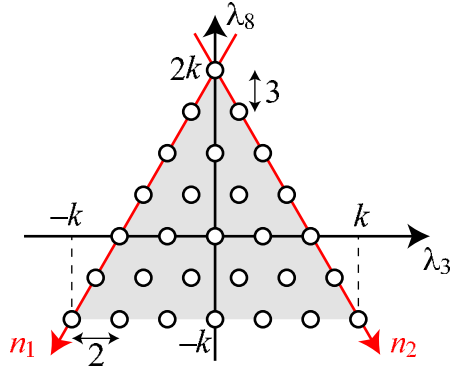}}
\caption{The $(k,k)$ representation on the weight space.
This example shows $(6,6)$.}
\label{trig.eps}
\end{figure}

The $SU(3)_V$ representation $(k,k)$ contains weight vectors
corresponding to lattice points
in a triangle as shown in Figure \ref{trig.eps}.
The multiplicity for every weight is one.
We label these weights by two integers
$n_1$ and $n_2$ satisfying
\begin{equation}
n_1,n_2\geq 0,\quad
n_1+n_2\leq k.
\end{equation}
The eigenvalues of Cartan generators $\lambda_3$ and $\lambda_8$ are
\begin{equation}
\mu(\lambda_3)=n_2-n_1,\quad
\mu(\lambda_8)=2k-n_1-n_2,
\end{equation}
and the character of this representation is
\begin{equation}
\chi_{(k,k)}(y_3,y_8)=\sum_{\mu\in(k,k)}
y_3^{\mu(\lambda_3)}
y_8^{\mu(\lambda_8)}
=\sum_{n_1=0}^k\sum_{n_2=0}^{k-n_1}y_3^{n_2-n_1}y_8^{2k-n_1-n_2}.
\end{equation}
If we introduce $n_3=k-n_1-n_2$,
we can rewrite the product over representations
$(k,k)$ ($k=0,1,2,\ldots$) and weights $\mu\in(k,k)$ as
the product over three independent
non-negative integers $n_1$, $n_2$, and $n_3$,
\begin{equation}
\prod_{k=0}^\infty\prod_{\mu\in(k,k)}(\cdots)
=\prod_{n_1=0}^\infty\prod_{n_2=0}^\infty\prod_{n_3=0}^\infty(\cdots).
\end{equation}
By using this, we can rewrite the infinite product appearing in the modified
plethystic exponential
as
\begin{equation}
\PE'
\sum_{k=0}r x^k\chi_{(k,k)}(y_3,y_8)
=\prod_{k=0}^\infty\prod_{\mu\in(k,k)}
\frac{\sqrt{rx^ky_3^{\mu(\lambda_3)}y_8^{\mu(\lambda_8)}}}
{1-rx^ky_3^{\mu(\lambda_3)}y_8^{\mu(\lambda_8)}}
=\frac{1}{F(r;{\bm p})},
\end{equation}
where we defined the function
\begin{equation}
F(r;{\bm p})=
\prod_{n_1,n_2,n_3=0}^\infty
\frac{1-rp_1^{n_1}p_2^{n_2}p_3^{n_3}}{\sqrt{rp_1^{n_1}p_2^{n_2}p_3^{n_3}}},
\end{equation}
with $\bm p=(p_1,p_2,p_3)$ given by
\begin{equation}
p_1=\frac{x}{y_3y_8},\quad
p_2=\frac{xy_3}{y_8},\quad
p_3=xy_8^2.
\label{p123}
\end{equation}
With this function we can express the modified
plethystic exponential of ${\cal I}_{\rm sp}^{\rm vector}$ as
\begin{equation}
\PE'{\cal I}^{\rm vector}_{\rm sp}=
\prod_{\alpha\in{\rm root}}
F(q^{-i\alpha(\sigma)};{\bm p})
F(q^{-i\alpha(\sigma)}x^3;{\bm p}).
\end{equation}

\subsection{Hypermultiplets}
A 6d hypermultiplet contains four real scalar fields $q_i$ ($i=1,2,3,4$)
and a symplectic Majorana-Weyl spinor field $\psi$ as on-shell
degrees of freedom.
It is difficult to give the off-shell
supersymmetry transformation of hypermultiplets
for full ${\cal N}=(1,0)$ supersymmetry.
Instead, we give the off-shell supersymmetry only for $\cal Q$ used in localization.
This is easily done
by following the prescription in \cite{Hosomichi:2012ek}
used for 5d hypermultiplets.
In addition to the Killing spinor $\varepsilon$,
we introduce another spinor $\check\varepsilon$ with negative chirality
satisfying
\begin{equation}
\varepsilon_I\check\varepsilon_J=0,\quad
(\varepsilon\Gamma^M\varepsilon)
+(\check\varepsilon\Gamma^M\check\varepsilon)=0.
\label{checkcond}
\end{equation}
The transformation laws with the Grassmann odd parameters
$(\epsilon,\check\epsilon)=(\eta\varepsilon,\eta\check\varepsilon)$
are
\begin{eqnarray}
\delta q_i&=&i(\epsilon\rho_i\psi),\nonumber\\
\delta\psi&=&i\ol\rho_i\Gamma^M\epsilon D_M q_i
+\ol\rho_i\check\epsilon f_i
+4i\ol\rho_i\kappa q_i,\nonumber\\
\delta f_i&=&
(\check\epsilon\rho_i\Gamma^MD_M\psi)
+(\check\epsilon\rho_i\ol\rho_j\lambda)q_j,
\end{eqnarray}
where $f_i$ ($i=1,2,3,4$) are real auxiliary fields.
$\rho_i$ and $\ol\rho_i$ ($i=1,2,3,4$)
defined by
\begin{equation}
\rho_i=(\sigma_1,\sigma_2,\sigma_3,-i\bm1_2),\quad
\ol\rho_i=(\sigma_1,\sigma_2,\sigma_3,i\bm1_2)
\end{equation}
are $SU(2)_F\times SU(2)_R$ invariant tensors,
where $SU(2)_F$ is the flavor symmetry acting on the scalar fields in the
hypermultiplet.
The conditions in (\ref{checkcond})
guarantee the closure of the algebra
and the supersymmetry invariance of the action
\begin{equation}
{\cal L}_{\rm hyper}=D_Mq_iD^Mq_i
+\psi\Gamma^MD_M\psi
+(\tau_a)_{ij}q_i[D_a,q_j]
+2\psi\ol\rho_i[\lambda,q_i]
+\frac{1}{5}Rq_iq_i
+f_if_i,
\label{hyperkin}
\end{equation}
where $(\tau_a)_{ij}\equiv-(1/2)\tr(\tau_a\rho_i\ol\rho_j)$ is the 't Hooft symbol
and $R$ is the scalar curvature of the background manifold.

On $\bm S^5\times\RR$,
we adopt the following $\check\varepsilon$
and its $\star$ conjugate
\begin{equation}
\check\varepsilon_1
=\left(\begin{array}{c}
0 \\
\vdots \\
0 \\
0 \\
ie^{-\frac{t}{2}}
\end{array}\right),\quad
\check\varepsilon_2
=\left(\begin{array}{c}
0 \\
\vdots \\
0 \\
-ie^{+\frac{t}{2}} \\
0
\end{array}\right),\quad
\check\varepsilon^\star=\check\varepsilon^\dagger|_{t\rightarrow -t}.
\end{equation}
The action (\ref{hyperkin})
is in fact not only ${\cal Q}$-invariant
but also ${\cal Q}$-exact
on $\bm S^5\times\RR$, and can be written as
\begin{equation}
{\cal L}_{\rm hyper}=\frac{1}{2}{\cal Q}[({\cal Q}\psi)^\star\psi].
\end{equation}
We use this action for the computation of the spectrum.

We assume that the hypermultiplet belongs to
$R+\ol R$ of the gauge group.
We denote a weight of $R$ by $\rho$,
and then the weight of its anti-particle is $-\rho\in\ol R$.
If we specify a weight vector $\rho\in R$ and
an $SU(3)_V$ weight vector $\mu\in(k,m)$,
we obtain the spectrum of $q_i$ excitations shown in Table \ref{table:qspec},
and $\psi$ excitations in Table \ref{table:psispec}.
\begin{table}[htb]
\caption{The spectrum of the scalar fields $q_i$. $\rho$ stands for $\rho(\sigma)$.}
\label{table:qspec}
\begin{center}
\begin{tabular}{ccccclc}
\hline
\hline
ID & $H$ & $Q_V$ & $\tau_3$ & Range of $m$ & $H-Q_V-2\tau_3$ & $Q_V+\frac{3}{2}\tau_3$ \\
\hline
$[q1]$ & $-i\rho+k+2$ & $m$ & $-1$ & $-k\leq m\leq k$ & $-i\rho+k-m+4$ & $m-\frac{3}{2}$ \\
$[q2]$ & $-i\rho+k+2$ & $m$ & $+1$ & $-k\leq m\leq k$ & $-i\rho+k-m$ & $m+\frac{3}{2}$ \\
$[q3]$ & $+i\rho+k+2$ & $m$ & $-1$ & $-k\leq m\leq k$ & $+i\rho+k-m+4$ & $m-\frac{3}{2}$ \\
$[q4]$ & $+i\rho+k+2$ & $m$ & $+1$ & $-k\leq m\leq k$ & $+i\rho+k-m$ & $m+\frac{3}{2}$ \\
\hline
\end{tabular}
\end{center}
\end{table}
\begin{table}[htb]
\caption{The spectrum of $\psi$. $\rho$ stands for $\rho(\sigma)$.}
\label{table:psispec}
\begin{center}
\begin{tabular}{cccclc}
\hline
\hline
ID & $H$ & $Q_V$ & Range of $m$ & $H-Q_V-2\tau_3$ & $Q_V+\frac{3}{2}\tau_3$ \\
\hline
$[\psi1]$ & $-i\rho+k+\frac{5}{2}$ & $m-\frac{3}{2}$ & $-k\leq m\leq k$ & $-i\rho+k-m+4$ & $m-\frac{3}{2}$ \\
$[\psi2]$ & $-i\rho+k+\frac{3}{2}$ & $m+\frac{3}{2}$ & $-k\leq m\leq k-2$ & $-i\rho+k-m$ & $m+\frac{3}{2}$ \\
$[\psi3]$ & $+i\rho+k+\frac{5}{2}$ & $m-\frac{3}{2}$ & $-k\leq m\leq k$ & $+i\rho+k-m+4$ & $m-\frac{3}{2}$ \\
$[\psi4]$ & $+i\rho+k+\frac{3}{2}$ & $m+\frac{3}{2}$ & $-k\leq m\leq k-2$ & $+i\rho+k-m$ & $m+\frac{3}{2}$ \\
\hline
\end{tabular}
\end{center}
\end{table}

After the cancellation, only bosonic modes $[q2]$ and $[q4]$ with $m=k$ contribute to
the single-particle index by
\begin{equation}
{\cal I}^{\rm hyper}_{\rm sp}(q,x,y_3,y_8)
=\sum_{\rho\in R}\sum_{k=0}^\infty
\left(q^{-i\rho(\sigma)}x^{k+\frac{3}{2}}
+q^{+i\rho(\sigma)}x^{k+\frac{3}{2}}\right)
\chi_{(k,k)}(y_3,y_8),
\end{equation}
and the modified plethystic exponential of this is
\begin{equation}
\PE'{\cal I}_{\rm sp}^{\rm hyper}=
\frac{1}{
\prod_{\rho\in R}F(q^{-i\rho(\sigma)}x^{\frac{3}{2}};{\bm p})
F(q^{+i\rho(\sigma)}x^{\frac{3}{2}};{\bm p})}.
\end{equation}

\section{Partition function from 6d index}\label{limit.sec}
\subsection{One-loop partition function}
By collecting all the contributions, we obtain the total index
\begin{align}
{\cal I}(x,y_3,y_8)&=
\left(C_0\PE'{\cal I}_{\rm sp}^{\rm Cartan}\right)
\left(\prod_{i=1}^{\rank G}\oint_0^{\beta/2\pi}d\sigma_i\right) Z^{\rm(6d)}_{\rm 1-loop},\nonumber\\
Z^{\rm(6d)}_{\rm 1-loop}&=\frac{\prod_{\alpha\in{\rm root}}F(q^{-i\alpha(\sigma)};{\bm p})F(q^{-i\alpha(\sigma)}x^3;{\bm p})}
     {\prod_{\rho\in R}F(q^{-i\rho(\sigma)}x^{\frac{3}{2}};{\bm p})F(q^{+i\rho(\sigma)}x^{\frac{3}{2}};{\bm p})}.
\label{d6index}
\end{align}
Precisely, we should include the factor $e^{-S_0}$ that comes from
the classical action $S_0$ at the saddle points.
Such a contribution exists in 5d theory, and if we
could lift the action of the 5d theory to 6d,
the classical 6d action should also contribute to the index.
However, we do not include it here because
we do not know the 6d lift of
a general 5d theory.
We will introduce the classical contribution after
taking the small radius limit.

To take the small radius limit,
let us relate the fugacities $q$, $x$, $y_3$ and $y_8$
to the geometry of $\bm S^5\times\bm S^1$.
If we set
\begin{equation}
x=q^{1+iw_0},\quad
y_3=q^{iw_3},\quad
y_8=q^{iw_8},
\label{eq74}
\end{equation}
and $q=e^{-\beta}$,
the twist operator (\ref{theopo}) becomes
\begin{equation}
{\cal O}=q^{H+iw_0Q_V+iw_3\lambda_3+iw_8\lambda_8}q^{(-\frac{1}{2}+\frac{3i}{2}w_0)\tau_3}.
\end{equation}
Two factors on the right hand side correspond to geometric and internal symmetries.
The geometric factor represents the shift
generated by
$H+iw_0Q_V+iw_3\lambda_3+iw_8\lambda_8$,
and the trace in the index
(\ref{indexdef})
corresponds to the
identification
(\ref{ident}) with the squashing parameters
\begin{equation}
\phi_1=w_0-w_3-w_8,\quad
\phi_2=w_0+w_3-w_8,\quad
\phi_3=w_0+2w_8.
\label{phi123}
\end{equation}
We can read off the relations
\begin{equation}
p_i=q^{\omega_i},\quad
\omega_i=1+i\phi_i.
\end{equation}
from (\ref{p123}), (\ref{eq74}), and (\ref{phi123}).
The partition function on the squashed sphere
is given by taking the small radius limit $\beta\rightarrow 0$;
\begin{equation}
Z_{\rm pert}=\lim_{q\rightarrow 1}{\cal I}(x=q^{1+iw_0},y_3=q^{iw_3},y_8=q^{iw_8}).
\end{equation}
In this limit the function $F$ reduces to
\begin{equation}
F(q^c;{\bm p})\rightarrow
\prod_{n_1,n_2,n_3=0}^\infty\left(c+n_1\omega_1+n_2\omega_2+n_3\omega_3\right)
=\frac{1}{\Gamma_3(c,{\bm\omega})},
\label{flimitgamma}
\end{equation}
and
we obtain the one-loop determinant of the 5d theory
\begin{equation}
Z^{(6d)}_{\rm 1-loop}
\rightarrow
Z_{\rm 1-loop}
=
\frac
{
\prod_{\alpha\in{\rm root}}
S_3(-i\alpha(\sigma),{\bm\omega})
}
{
\prod_{\rho\in R}
S_3(-i\rho(\sigma)+\frac{\omega_{\rm tot}}{2},{\bm\omega})
}.
\label{zoneloop}
\end{equation}
$\Gamma_3(z,\bm\omega)$ and
$S_3(z,\bm\omega)$ are the triple gamma function and the triple sine function,
respectively.
In general, the multiple gamma function $\Gamma_r(z,\bm\omega)$ \cite{Barnes}
and
the multiple sine function $S_r(z,\bm\omega)$ \cite{Kurokawa1,Kurokawa2,Kurokawa3,KurokawaKoyama}
depend on the $r$-component period $\bm\omega=\{\omega_1,\ldots,\omega_r\}$,
and are defined by
\begin{eqnarray}
\Gamma_r(z,\bm\omega)
&=&\prod_{{\bm n}\geq 0}({\bm n}\cdot{\bm\omega}+z)^{-1},
\nonumber\\
S_r(z,\bm\omega)
&=&
\Gamma_r(z,\bm\omega)^{-1}
\Gamma_r(\omega_{\rm tot}-z,\bm\omega)^{(-1)^r}
\nonumber\\&=&
\left(\prod_{{\bm n}\geq 0}({\bm n}\cdot{\bm\omega}+z)\right)
\left(\prod_{{\bm n}\geq 1}({\bm n}\cdot{\bm\omega}-z)\right)^{(-1)^{r-1}},
\nonumber\\
\omega_{\rm tot}&=&\sum_{i=1}^r\omega_i.
\end{eqnarray}

The integration variables $\sigma_i$
originally take value in ${\bm S}^1$ with period $2\pi/\beta$.
In the small radius limit the ${\bm S}^1$ is replaced by $\RR$,
and in the 5d theory $\sigma_i$ are integrated over the real axis;
\begin{equation}
\prod_{i=1}^{\rank G}\oint_0^{2\pi/\beta}d\sigma_i
\rightarrow
\prod_{i=1}^{\rank G}\int_{-\infty}^{\infty}d\sigma_i.
\end{equation}

We also have the constant factor independent of $\sigma$:\footnote{
This factor does not affect
the $N^3$ behavior of the free energy of the 5d ${\cal N}=2$ supersymmetric
Yang-Mills theory studied in \cite{Kallen:2012zn}.
However, it depends on the squashing parameters
$\phi_i$ and would have physical significance at finite $N$.
This factor is also given in \cite{Spiridonov:2012de}.}
\begin{equation}
C({\bm\omega})=\lim_{\beta\rightarrow 0}C_0\PE'{\cal I}_{\rm sp}^{\rm Cartan}
=\frac{1}{|W|}\left(\frac{S_3'(0;{\bm\omega})}{2\pi}\right)^{\rank G},
\label{comega}
\end{equation}
where $S'_3(z;{\bm\omega})=\partial S_3(z;{\bm\omega})/\partial z$.

We have obtained the perturbative partition function
(\ref{final}) except the exponential factor comming from the classical action.
For $SU(3)\times U(1)$ symmetric squashed ${\bm S}^5$,
the classical action has already been computed
in \cite{Imamura:2012xg} and shown to proportional to the prepotential.
In the following we will first confirm that
for such homogeneous squashed spheres the classical action indeed gives
the exponential factor in (\ref{final}).
For a generic squashed sphere, unfortunately,
we do not know the precise form of the classical action,
and we cannot prove (\ref{final}).
In this case we will ``define'' the prepotential
so that it is proportional to
the classical action,
and
determine the proportionality
constant by looking at the asymptotic behavior of the
one-loop factor.

\subsection{${\cal N}=1/4$ case}\label{n14.sec}
In this and the next subsections,
we consider homogeneous squashed ${\bm S}^5$.
Generically, a squashed ${\bm S}^5$ has isometry
$U(1)^3\subset SU(4)_{\rm iso}$, and is not homogeneous.
We obtain a homogeneous squashed sphere
when the operator ${\cal O}$ in (\ref{theopo})
preserves $SU(3)\times U(1)$ isometry,
which may differ from $SU(3)_V\times U(1)_V$ preserved by $I_{\CC^3}$.
There are two essentially different cases.
One is
\begin{equation}
\phi_1=\phi_2=\phi_3=-u.
\end{equation}
In this case,
the compactification preserves $SU(3)_V\times U(1)_V$, and
the supercharge $\cal Q$ is $SU(3)_V$-singlet.
This corresponds to ${\cal N}=1/4$ theory constructed in
\cite{Imamura:2012xg}.
The periods are
\begin{equation}
{\bm\omega}=(1-iu){\bm1},\quad
{\bm1}=(1,1,1).
\end{equation}
The classical action at a saddle point $\sigma$ is \cite{Imamura:2012xg}%
\footnote{The integration variable $a$ in \cite{Imamura:2012xg}
is related to $\sigma$ by $a=\sigma/(1+u^2)^{1/2}$.}
\begin{equation}
S_{{\cal N}=\frac{1}{4}}=\frac{(2\pi)^3}{(1-iu)^3}{\cal F}(\sigma),
\label{n14classical}
\end{equation}
and this gives the exponential factor in (\ref{final}).

When we brought (\ref{n14classical}) from \cite{Imamura:2012xg},
we changed the convention for the prepotential.
The prepotential is a cubic function of $\sigma$,
and is expanded schematically as
\begin{equation}
{\cal F}(\sigma)=g_3\sigma^2+g_2\sigma^2+g_1\sigma+g_0.
\end{equation}
The coefficients $g_i$ ($i=0,1,2,3$) are coupling constants of the theory,
and if the background is flat, there is standard normalization of them.
In a curved manifold,
however,
coefficients in the Lagrangian density
may depend on the coordinates, and
there is no standard choice of the normalization of $g_i$,
except $g_3$, which is proportional to the Chern-Simons level and
quantized.
Due to this ambiguity in the definition of $g_i$ ($i=0,1,2$),
the prepotential depends on the convention.
In particular, by the rescaling of the coefficients
$g_i\rightarrow g_i'=c^{3-i}g_i$,
the prepotential changes as
\begin{equation}
{\cal F}(\sigma)
\rightarrow
{\cal F}'(\sigma)
=c^3{\cal F}\left(\frac{\sigma}{c}\right).
\label{fcfp}
\end{equation}
If we denote the prepotential in
\cite{Imamura:2012xg} by ${\cal F}'$,
it is related to ${\cal F}$ here
in this way with $c=(1+u^2)^{-1/2}$.

As is argued in \cite{Imamura:2012xg},
the perturbative partition function
of ${\cal N}=1/4$ theory is independent of the squashing parameter $u$.
This is shown as follows.
By introducing new integration variable $\sigma'=\sigma/(1-iu)$
and using the scaling property of the multiple sine function
$S_r(cz,c\bm\omega)
=S_r(z,\bm\omega)$,
we can eliminate the $u$-dependence
from the one-loop factor and
prefactor $C({\bm\omega})$ combined with the integration measure:
\begin{align}
&Z_{\rm 1-loop}
=
\frac
{
\prod_{\alpha\in{\rm adj}}
S_3(-i\alpha(\sigma'),{\bm1})
}
{
\prod_{\rho\in R}
S_3(-i\rho(\sigma')+\frac{3}{2},{\bm1})
},\nonumber\\
&C({\bm\omega})\int\prod_{i=1}^{\rank G} d\sigma_i
=C({\bm1})
\int\prod_{i=1}^{\rank G}d\sigma'_i.
\label{zoneloop2}
\end{align}
We can also remove $u$ from the classical action
by changing the convention of the prepotential
by (\ref{fcfp}) with $c=1-iu$:
\begin{equation}
S_{{\cal N}=\frac{1}{4}}=(2\pi)^3
{\cal F}(\sigma').
\label{n14classical2}
\end{equation}

\subsection{${\cal N}=3/4$ case}\label{n34.sec}
The other choice of the squashing parameter preserving an
$SU(3)\times U(1)$ isometry
is
\begin{equation}
\phi_1=\phi_2=-\phi_3=u.
\end{equation}
Let $SU(3)_{\rm iso}\times U(1)_{\rm iso}$ be the preserved isometry.
In this case, $SU(3)_{\rm iso}$ is different from $SU(3)_V$,
and the supercharge ${\cal Q}$ belongs to an $SU(3)_{\rm iso}$ triplet.
This gives ${\cal N}=3/4$ theory in \cite{Imamura:2012xg}.
The periods are
\begin{equation}
\omega_1=1+iu,\quad
\omega_2=1+iu,\quad
\omega_3=1-iu.
\end{equation}
The classical action at a saddle point $\sigma$ is \cite{Imamura:2012xg}
\begin{equation}
S_{{\cal N}=\frac{3}{4}}=\frac{(2\pi)^3}{(1+iu)^2(1-iu)}{\cal F}(\sigma).
\end{equation}
(We took account of the difference in the convention for the prepotential
as in \S\ref{n14.sec}.)
This gives the exponential factor in (\ref{final}).
In this case, the partition function depends on the squashing parameter $u$
in a non-trivial way.

As is pointed out in \cite{Jafferis:2012iv},
the asymptotic behavior of the one-loop partition function
on the round $\bm S^5$
can be interpreted as the quantum correction
to Chern-Simons levels.
This is also the case in ${\cal N}=1/4$ theories because
the classical action and the one-loop partition function
are independent of the squashing parameter $u$.
In ${\cal N}=3/4$ case, they non-trivially depends on
$u$.
Let us conform that the
asymptotic behavior of $Z_{\rm 1-loop}$ is still consistent to the quantization
of Chern-Simons levels.
For simplicity, let us consider a $U(1)$ gauge theory with
the Chern-Simons term
\begin{equation}
{\cal L}=\frac{k}{6(2\pi)^2}\int A\wedge F\wedge F.
\end{equation}
The prepotential contains the corresponding term
\begin{equation}
{\cal F}(a)=\frac{k}{6(2\pi)^2}a^3,
\end{equation}
and this contributes to the
classical action by
\begin{equation}
S_{{\cal N}=\frac{3}{4}}=\frac{\pi k}{3(1+iu)^2(1-iu)}\sigma^3.
\label{sn34}
\end{equation}

We compare this to the asymptotic value of the one-loop partition function
(\ref{zoneloop}).
We consider the contribution of a single hypermultiplet
with charge $1$.
This gives the factor $1/S_3(\frac{\omega_{\rm tot}}{2}-i\sigma;{\bm\omega})$
in the one-loop determinant.
In the asymptotic region $|\Re\sigma|\rightarrow \infty$,
the triple sine function behaves as\footnote{This can be obtained
from some formulae of the multiple gamma function in \cite{Ruijsenaars}}
\begin{equation}
\log S_3\left(\frac{\omega_{\rm tot}}{2}-i\sigma,{\bm\omega}\right)
\sim \sign(\sigma)\left(-\frac{\pi}{6\omega_1\omega_2\omega_3}\sigma^3
-\frac{\pi(\omega_1^2+\omega_2^2+\omega_3^2)}{24\omega_1\omega_2\omega_3}\sigma\right).
\end{equation}
(We assumed $\Re\omega_i>0$.)
The coefficient of the cubic term is
\begin{equation}
-\sign(\sigma)\frac{\pi}{6\omega_1\omega_2\omega_3}
=-\sign(\sigma)\frac{\pi}{6(1+iu)^2(1-iu)}.
\end{equation}
Comparing this to
(\ref{sn34})
we find that this effectively shifts the coefficient $k$
by $\pm 1/2$.
This is the same as the one-loop correction
to the Chern-Simons level by
the fermion in the single hypermultiplet.

\subsection{General squashed $\bm S^5$}
The supersymmetric action for a general squashed
$\bm S^5$ has not been explicitly obtained,
and thus we cannot directly determine the
classical action $S_0(\sigma)$.
However,
by taking advantage of the ambiguity in the
definition of the prepotential that we mentioned in \S\ref{n14.sec},
we define the prepotential
so that it is proportional to the classical action at
saddle points.
Then, what we need to do
is to determine the proportionality constant.
For this purpose we can use the relation between
the classical action and the asymptotic form
of the one-loop determinant as discussed in the previous subsection.
We obtain
\begin{equation}
S_0=\frac{(2\pi)^3}{\omega_1\omega_2\omega_3}{\cal F}(\sigma)
\end{equation}
for general squashed ${\bm S}^5$,
and this gives the exponential factor in (\ref{final}).

\section{Discussion}
We computed the perturbative partition function
of 5d ${\cal N}=1$ supersymmetric gauge theories
defined on a squashed ${\bm S}^5$.
The result is expressed in a simple form
containing the triple sine function.
This is analogous to the ${\bm S}^3$ partition function
written in terms of the double sine function.
The ${\bm S}^5$ partition function depends on the
squashing parameters $(\phi_1,\phi_2,\phi_3)$ through the periods
$\omega_i=1+i\phi_i$.

To determine the classical action
at saddle points for general squashing parameters,
we used the convention of the prepotential in which the
classical action at saddle points is proportional to the prepotential,
and the proportionality constant is
determined by
the consistency of the asymptotic form of the one-loop determinant factor
to the quantization of Chern-Simons levels.
We do not know the explicit form of the action
on a general squashed $\bm S^5$, and
the precise meaning of the coefficients in the prepotential
in the convention is not clear.
In general, we cannot obtain Chern-Simons terms in 5d by the dimensional
reduction from 6d,
we need to rely on the direct analysis in 5d such as Noether procedure
to construct the supersymmetric action in general squashed ${\bm S}^5$.

An important application of the partition function
is the confirmation
of predictions of AdS/CFT.
For the round $\bm S^5$, the large $N$ partition function
of $USp(2N)$ supersymmetric QCD with
$N_f\leq 7$ flavors is computed in \cite{Jafferis:2012iv},
and precise agreement of the large $N$ leading term to the
prediction of the gravity dual is confirmed.
It is interesting to extend this analysis to squashed $\bm S^5$.
Another interesting theory is the $U(N)$ supersymmetric Yang-Mills theory
with a single adjoint hypermultiplet,
which is related to
the $(2,0)$ theory realized on M5-branes via $\bm S^1$ compactification.
The $N^3$ behavior of this partition function is confirmed in \cite{Kim:2012av,Kallen:2012zn}.
It is important to
establish the relation between the gauge theory
and the gravity dual
for an arbitrary mass parameter $\Delta$ and squashing parameters $\phi_i$.

\section*{Acknowledgments}
Y.~I. is partially supported by Grant-in-Aid for Scientific Research
(C) (No.24540260), Ministry of Education, Science and Culture, Japan.

\appendix
\section{Appendix}
\subsection{Dirac matrices}

We use the following representation of the 5d Dirac matrices
\begin{equation}
\gamma_i=\left(\begin{array}{cc} 0 & \sigma_i \\ \sigma_i & 0 \end{array}\right),\quad
\gamma_4=\left(\begin{array}{cc} 0 & -i{\bm1}_2 \\ i{\bm1}_2 & 0 \end{array}\right),\quad
\gamma_5=\left(\begin{array}{cc} \bm1_2 & 0 \\ 0 & -\bm1_2 \end{array}\right).
\end{equation}
where $\sigma_i$ ($i=1,2,3$) are the Pauli matrices
\begin{equation}
\sigma_1=\left(\begin{array}{cc} 0 & 1 \\ 1 & 0 \end{array}\right),\quad
\sigma_2=\left(\begin{array}{cc} 0 & -i \\ i & 0 \end{array}\right),\quad
\sigma_3=\left(\begin{array}{cc} 1 & 0 \\ 0 & -1 \end{array}\right).
\end{equation}
The 6d Dirac matrices are
\begin{equation}
\Gamma_\mu=\left(\begin{array}{cc} 0 & \gamma_\mu \\ \gamma_\mu & 0 \end{array}\right),\quad
\Gamma_6=\left(\begin{array}{cc} 0 & -i\bm1_4 \\ i\bm1_4 & 0 \end{array}\right),\quad
\Gamma_7=\left(\begin{array}{cc} \bm1_4 & 0 \\ 0 & -\bm1_4 \end{array}\right).
\end{equation}
The bilinear of $SU(2)$ doublet spinors is defined by
\begin{equation}
\psi\chi=\epsilon^{IJ}C^{ab}\psi_{Jb}\chi_{Ia}
\end{equation}
where $\epsilon^{IJ}$ is $SU(2)$ invariant anti-symmetric tensor
with component $\epsilon^{12}=-\epsilon^{21}=1$,
and $C^{ab}$ is the charge conjugation matrix with components
\begin{equation}
C^{ab}=\left(\begin{array}{cccc}
&& i\sigma_2 \\
&&& -i\sigma_2 \\
 i\sigma_2 \\
& -i\sigma_2
\end{array}\right).
\end{equation}
The antisymmetric tensor is defined by
\begin{equation}
\Gamma^{M_1\cdots M_6}\Gamma^7
=\epsilon^{M_1\cdots M_6}{\bm1}_8.
\end{equation}


\begin{thebibliography}{99}
\bibitem{Pestun:2007rz} 
  V.~Pestun,
  ``Localization of gauge theory on a four-sphere and supersymmetric Wilson loops,''
  Commun.\ Math.\ Phys.\  {\bf 313}, 71 (2012)  [arXiv:0712.2824 [hep-th]].
\bibitem{Alday:2009aq} 
  L.~F.~Alday, D.~Gaiotto and Y.~Tachikawa,
  ``Liouville Correlation Functions from Four-dimensional Gauge Theories,''
  Lett.\ Math.\ Phys.\  {\bf 91}, 167 (2010)  [arXiv:0906.3219 [hep-th]].
\bibitem{Kapustin:2009kz} 
  A.~Kapustin, B.~Willett and I.~Yaakov,
  ``Exact Results for Wilson Loops in Superconformal Chern-Simons Theories with Matter,''
  JHEP {\bf 1003}, 089 (2010)  [arXiv:0909.4559 [hep-th]].
\bibitem{Jafferis:2010un} 
  D.~L.~Jafferis,
  JHEP {\bf 1205}, 159 (2012)
  [arXiv:1012.3210 [hep-th]].
\bibitem{Hama:2010av} 
  N.~Hama, K.~Hosomichi and S.~Lee,
  JHEP {\bf 1103}, 127 (2011)
  [arXiv:1012.3512 [hep-th]].
\bibitem{Drukker:2010nc} 
  N.~Drukker, M.~Marino and P.~Putrov,
  ``From weak to strong coupling in ABJM theory,''
  Commun.\ Math.\ Phys.\  {\bf 306}, 511 (2011)  [arXiv:1007.3837 [hep-th]].
\bibitem{Herzog:2010hf} 
  C.~P.~Herzog, I.~R.~Klebanov, S.~S.~Pufu and T.~Tesileanu,
  ``Multi-Matrix Models and Tri-Sasaki Einstein Spaces,''
  Phys.\ Rev.\ D {\bf 83}, 046001 (2011)  [arXiv:1011.5487 [hep-th]].
\bibitem{Marino:2011eh} 
  M.~Marino and P.~Putrov,
  ``ABJM theory as a Fermi gas,''
  J.\ Stat.\ Mech.\  {\bf 1203}, P03001 (2012)  [arXiv:1110.4066 [hep-th]].

\bibitem{Kallen:2012cs} 
  J.~Kallen and M.~Zabzine,
  ``Twisted supersymmetric 5D Yang-Mills theory and contact geometry,''
  JHEP {\bf 1205}, 125 (2012)
  [arXiv:1202.1956 [hep-th]].
\bibitem{Kallen:2012va} 
  J.~Kallen, J.~Qiu and M.~Zabzine,
  ``The perturbative partition function of supersymmetric 5D Yang-Mills theory with matter on the five-sphere,''
  arXiv:1206.6008 [hep-th].
\bibitem{Kim:2012av} 
  H.~-C.~Kim and S.~Kim,
  ``M5-branes from gauge theories on the 5-sphere,''
  arXiv:1206.6339 [hep-th].
\bibitem{Seiberg:1996bd} 
  N.~Seiberg,
  ``Five-dimensional SUSY field theories, nontrivial fixed points and string dynamics,''
  Phys.\ Lett.\ B {\bf 388}, 753 (1996)
  [hep-th/9608111].
\bibitem{Douglas:2010iu} 
  M.~R.~Douglas,
  ``On D=5 super Yang-Mills theory and (2,0) theory,''
  JHEP {\bf 1102}, 011 (2011)  [arXiv:1012.2880 [hep-th]].
\bibitem{Lambert:2010iw} 
  N.~Lambert, C.~Papageorgakis and M.~Schmidt-Sommerfeld,
  ``M5-Branes, D4-Branes and Quantum 5D super-Yang-Mills,''
  JHEP {\bf 1101}, 083 (2011)  [arXiv:1012.2882 [hep-th]].
\bibitem{Hosomichi:2012ek} 
  K.~Hosomichi, R.~-K.~Seong and S.~Terashima,
  ``Supersymmetric Gauge Theories on the Five-Sphere,''
  arXiv:1203.0371 [hep-th].
\bibitem{Kawano:2012up} 
  T.~Kawano and N.~Matsumiya,
  ``5D SYM on 3D Sphere and 2D YM,''
  arXiv:1206.5966 [hep-th].
\bibitem{Kim:2012gu} 
  H.~-C.~Kim, S.~-S.~Kim and K.~Lee,
  ``5-dim Superconformal Index with Enhanced En Global Symmetry,''
  arXiv:1206.6781 [hep-th].
\bibitem{Terashima:2012ra} 
  S.~Terashima,
  ``On Supersymmetric Gauge Theories on S$^4$ x S$^1$,''
  arXiv:1207.2163 [hep-th].
\bibitem{Kim:2012tr} 
  H.~-C.~Kim and K.~Lee,
  ``Supersymmetric M5 Brane Theories on R x CP2,''
  arXiv:1210.0853 [hep-th].
\bibitem{Fukuda:2012jr} 
  Y.~Fukuda, T.~Kawano and N.~Matsumiya,
  ``5D SYM and 2D q-Deformed YM,''
  arXiv:1210.2855 [hep-th].
\bibitem{Hama:2011ea} 
  N.~Hama, K.~Hosomichi and S.~Lee,
  ``SUSY Gauge Theories on Squashed Three-Spheres,''
  JHEP {\bf 1105}, 014 (2011)  [arXiv:1102.4716 [hep-th]].
\bibitem{Imamura:2011wg} 
  Y.~Imamura and D.~Yokoyama,
  ``N=2 supersymmetric theories on squashed three-sphere,''
  Phys.\ Rev.\ D {\bf 85}, 025015 (2012)
  [arXiv:1109.4734 [hep-th]].
\bibitem{Hama:2012bg} 
  N.~Hama and K.~Hosomichi,
  ``Seiberg-Witten Theories on Ellipsoids,''
  JHEP {\bf 1209}, 033 (2012)
  [arXiv:1206.6359 [hep-th]].
\bibitem{Dolan:2011rp} 
  F.~A.~H.~Dolan, V.~P.~Spiridonov and G.~S.~Vartanov,
  ``From 4d superconformal indices to 3d partition functions,''
  Phys.\ Lett.\ B {\bf 704}, 234 (2011)
  [arXiv:1104.1787 [hep-th]].
\bibitem{Gadde:2011ia}
  A.~Gadde and W.~Yan,
  ``Reducing the 4d Index to the $S^3$ Partition Function,''
  arXiv:1104.2592 [hep-th].
\bibitem{Imamura:2011uw}
  Y.~Imamura,
  ``Relation between the 4d superconformal index and the S$^3$ partition function,''
  [arXiv:1104.4482 [hep-th]].
\bibitem{Bhattacharya:2008zy} 
  J.~Bhattacharya, S.~Bhattacharyya, S.~Minwalla and S.~Raju,
  JHEP {\bf 0802}, 064 (2008)
  [arXiv:0801.1435 [hep-th]].
\bibitem{Pope:1980ub} 
  C.~N.~Pope,
  ``Eigenfunctions and Spin (c) structurs in CP**2,''
  Phys.\ Lett.\ B {\bf 97}, 417 (1980).
\bibitem{Barnes}
    E.~W.~Barnes,
    ``On the theory of the multiple gamma functions,''
    Trans. Cambridge Phil. Soc. {\bf19} (1904), 374
\bibitem{Kurokawa1}
    N.~Kurokawa,
    ``Multiple sine functions and Selberg zeta functions,''
    Proc. Japan Acad. A{\bf67} (1991), 61.
\bibitem{Kurokawa2}
    N.~Kurokawa,
    ``Gamma factors and Plancherel measures,''
    Proc. Japan Acad. A{\bf68} (1992), 256.
\bibitem{Kurokawa3}
    N.~Kurokawa,
    ``Multiple zeta functions: an example,''
    Adv. Stud. Pure Math. {\bf21} (1992) 219.
\bibitem{KurokawaKoyama}
    N.~Kurokawa and S.~Koyama,
    ``Multiple Sine Functions (I),''
    http://www.math.keio.ac.jp/library/research/report/2001/01005.pdf
\bibitem{Imamura:2012xg} 
  Y.~Imamura,
  ``Supersymmetric theories on squashed five-sphere,''
  arXiv:1209.0561 [hep-th].
\bibitem{Ruijsenaars}
  S.~N.~M.~Ruijsenaars,
  ``On Barnes' Multiple Zeta and Gamma Functions,''
  Advances in Mathematics {\bf156}, 107 (2000).
\bibitem{Jafferis:2012iv} 
  D.~L.~Jafferis and S.~S.~Pufu,
  ``Exact results for five-dimensional superconformal field theories with gravity duals,''
  arXiv:1207.4359 [hep-th].
\bibitem{Kallen:2012zn} 
  J.~Kallen, J.~A.~Minahan, A.~Nedelin and M.~Zabzine,
  ``$N^3$-behavior from 5D Yang-Mills theory,''  arXiv:1207.3763 [hep-th].  
\bibitem{LockhartVafa}
  G.~Lockhart, C.~Vafa,
  ``Superconformal Partition Functions and Non-perturbative Topological Strings,''
  arXiv:1210.5909 [hep-th]

\bibitem{Spiridonov:2012de} 
  V.~P.~Spiridonov,
  ``Modified elliptic gamma functions and 6d superconformal indices,''
  arXiv:1211.2703 [hep-th].
\end{thebibliography}
\end{document}